\documentclass[prd,superscriptaddress,nofootinbib,showkeys, preprint]{revtex4}
\usepackage{graphicx}
\usepackage[usenames,dvipsnames]{color}
\usepackage{amsmath,amssymb}
\usepackage{mathtools}
\usepackage[colorlinks]{hyperref}
\usepackage{float}

\newcommand{\ba}{\begin{eqnarray}}                                             
\newcommand{\ea}{\end{eqnarray}}

\newcommand{\be}{\begin{equation}}                                             
\newcommand{\ee}{\end{equation}}

\begin{document}
	\title{ Supersymmetry and Shape Invariance of exceptional orthogonal polynomials}
	\author{Satish Yadav}
	\email{s30mux@gmail.com}
		
		\affiliation{Department of Physics, Institute of Science, Banaras 
		Hindu University, 
		Varanasi - 221005, India}
		
		\author{Avinash Khare}
		\email{khare@physics.unipune.ac.in}
		\affiliation{Department of Physics, University of Pune, Ganeshkhind, Pune-411007, India}
		\author{Bhabani Prasad Mandal}
	\email{bhabani.mandal@gmail.com}
	\affiliation{Department of Physics, Institute of Science, Banaras 
		Hindu University, 
		Varanasi - 221005, India}
		\begin{abstract}
We discuss the exceptional Laguerre and the exceptional Jacobi orthogonal 
polynomials in the framework of the supersymmetric quantum mechanics (SUSYQM). 
We express the differential equations for the Jacobi and the Laguerre 
exceptional orthogonal polynomials (EOP) as the eigenvalue equations and make 
an analogy with the time independent Schr\"odinger equation to define 
``Hamiltonians" enables us to study the EOPs in the framework of the SUSYQM and to 
realize the
underlying shape invariance associated with such systems. We  show that the 
underlying shape invariance symmetry is responsible for the solubility of 
the differential equations associated with these polynomials. 		
\end{abstract}
	
	\keywords{Exceptional Orthogonal Polynomials, Rational extensions, Supersymmetric Quantum Mechanics, Shape Invariance}	
	\maketitle
	
	\section{Introduction}
	
In 2009 new families of orthogonal polynomials ( known as exceptional 
orthogonal polynomials (EOP)), related to some of the old classical orthogonal 
polynomials are discovered \cite{eop1,eop2,eop3}. Unlike the usual classical 
orthogonal polynomials, these EOPs start with degree $n= 1$ or higher integer 
values and still form a complete orthonormal set with respect to a positive 
definite inner product defined over a compact interval. Two of the well known 
classical orthogonal polynomials, namely the Laguerre orthogonal polynomials 
and the Jacobi orthogonal polynomials have been extended to the EOPs category.
By $X_m$ Laguerre (Jacobi) EOP means the complete set of  Laguerre (Jacobi) 
orthogonal polynomials with degree $\ge m$. Here $m$ is a positive integer and 
can have values $1,2,3 \cdots $ . 
Soon after this  remarkable discovery, the connection of EOPs with the 
translationally shape invariant potential was established 
\cite{re1,bqr,re2,re3,re4}. The  list of exactly solvable quantum mechanical 
potentials thus got enlarged with the corresponding energy eigen functions for 
the newly obtained exactly solvable systems being expressed in terms of EOPs.  
Such systems are known as rational extension of the original systems. These
developments have greatly helped in extending the list of the exactly solvable 
potentials  over the past decade 
\cite{re5,re6,re7,re8,re9,re10,re11,re12,re13,re131,re14,re15,ces,gt1,many1,
many2,many3,many4,ptt,pt1,pt2,pt3,pt4,ore1,ore11,ore3,ore81,qes6}. 	
Attempts have also been made to extend the 
classical Hermite polynomials to the EOP category \cite{hp}. 
	
In the present article we would like to study EOPs in the framework of 
the SUSYQM.The differential equations for  the Jacobi and the Laguerre EOPs can
be written as eigenvalue equations. We can associate a Hamiltonian operator to 
each of the EOPs by comparing the associated differential equation with the 
time independent Schr\"odinger equation, $ H\psi = E\psi $. Further, we show that
these Hamiltonians are factorizable in terms of the raising and the lowering 
operators. These raising (lowering) operators when act on the EOPs, increase 
(decrease) the order of the polynomials by one unit. This enables us to 
construct the supersymmetric partner Hamiltonians for  both the EOPs. 
Supersymmetric techniques are then used to get the solutions corresponding to 
these EOPs. Further by introducing an arbitrary integer parameter,  we 
construct a sequence of supersymmetric partner Hamiltonians  to show the 
underlying shape invariance associated with such systems.  For, simplicity we 
first present the  case of $X_1$ EOPs and later generalize to arbitrary m.
It is worth adding that a similar study has already been undertaken for the
classical orthogonal polynomials \cite{dgs,hkgm,bd} as well as for 
hypergeometric function \cite{dk}. 

The plan of the paper is as follows. In Sec. II we briefly introduce the basic 
ideas of SUSYQM and SI which will be necessary for the  analysis of the EOPs.
In section III we discuss $X_1$ Jacobi and the $X_1$ 
Laguerre EOPs within the framework of SUSYQM and bring out the SI 
associated with these EOPs which is responsible for the solubility
of these EOPs. In Sec IV we generalize these results to the $X_m$ EOPs. 
Finally we summarize the results in Sec V.
	 
\section{ Supersymmetry and Shape Invariance}
In this section we briefly review some key features of SUSYQM and shape 
invariance \cite{susy,susy1}. One dimensional Hamiltonian (in units of $2m=1,\,
\hbar =1$) is factorize in terms of first order operators as
\begin{eqnarray}
A &=& \frac{d}{dx} +W(x) \nonumber \\ 
A^\dagger  &=& -\frac{d}{dx} +W(x)	
\end{eqnarray}
where $W(x)$, a real function of x, is known as the super potential while  
$A^\dagger $ and $A$ are raising and lowering operators, analogous to case of 
simple harmonic oscillator. The supersymmetric partner Hamiltonians are then 
defined as
\begin{eqnarray}
H_- &=& A^\dagger A = -\frac{d^2}{dx^2} +V_-(x) \nonumber \\ 
H_+  &=& AA^\dagger= -\frac{d^2}{dx^2} +V_+(x)	
\end{eqnarray}	
Where $V_{\pm}= W^2\mp W^\prime (x)$.
These two Hamiltonians $H_{\mp}$ have exactly same spectrum except the lowest 
energy state of $H_-$, which is zero,i.e.
\begin{equation}
E_{0}^{(-)} = 0\,,~~E_{n}^{(+)} = E_{n+1}^{(-)}\,.
\end{equation}
The eigenfunctions of $H_\pm$ , \ $\psi^n_{\pm}, $ are 
interrelated through the operations of
$A^\dagger, A$. In particular,
\begin{equation}
\psi^{n}_{+} = A^{\dagger} \psi^{n+1}_{-}\,,~~\psi^{n+1}_{-} 
= A \psi^{n}_{+}\,,~~\psi^{0}_{-} = Ne^{\int^{x} W(y) dy}\,.
\end{equation}
Thus SUSYQM relates the bound state eigenvalues as well as the corresponding 
eigen functions of the partner Hamiltonians without actually finding them.

This analysis  can also be extended to the non-Hermitian systems\cite{nh}.In the case of non hermitian Hamiltonians the raising and lowering operators are not hermitian conjugate to each other and hence  the supersymmetry partner Hamiltonians  are like, $BA$ and $AB$ (instead of $A^\dagger A$ and $AA^\dagger$ ). In this case the spectrum
needs not be bounded from below. 

However, in case the two partner potentials are shape invariant, i.e. if  
they satisfy
\begin{equation}
V_{+}(x,a_0)=V_{-}(x,a_1) + R(a_1), 
\end{equation}
where $a_1$ is some function of $a_0$, i.e. $a_1 = f(a_0)$ and $R(a_1)$ is a
constant then one can show that the energy eigenvalues corresponding to  $V_+(x)$ are
given by
\begin{equation}
E_{n+1}^{(-)} = \sum_{i=1}^{n} R(a_i)\,,~~E_{0}^{(-)} = 0\,.
\end{equation}
If there is a family of $p$ shape invariant Hamiltonians, one can relate the $p-th$ eigenfunction of original Hamiltonian $H^1$ to $H^p$ as
\begin{equation}
    \psi_p^1(x;a_1)\propto B(x;a_1)B(x;a_2)...B(x;a_p)\psi_0^1(x;a_{p+1})
\end{equation}
where $B(x;a_p)$'s are the raising operators.
	
\section{$X_1$- Exceptional Orthogonal polynomials }

In this section, we discuss the $X_1$ Jacobi and the $X_1$ Laguerre 
polynomials within the framework of SUSYQM and SI techniques and bring out
the reasoning for the solubility of the corresponding $X_1$ Jacobi and
the $X_1$ Laguerre differential equations.

\subsection{$X_1$- Jacobi  Orthogonal polynomials}

The $X_1$-Jacobi orthogonal polynomials $\hat{P}_n^{(\alpha,\beta)}(x) $ satisfy the differential equation
\begin{equation}\label{eq13}
\begin{split}
\hat{P''}_n^{(\alpha,\beta)}(x)+\left(\frac{(\beta+\alpha+2)x-(\beta-\alpha)}
{x^2-1} + \frac{2(\beta-\alpha)}{(\beta+\alpha)-(\beta-\alpha)x}\right)
\hat{P'}_n^{(\alpha,\beta)}(x)\\ 
+\left(\frac{(\beta-\alpha)x-(n-1)(n+\alpha+\beta)}{x^2-1}+\frac{(\beta-\alpha)^2}{(\beta+\alpha)-(\beta-\alpha)x}\right)\hat{P}_n^{(\alpha,\beta)}(x)=0\\
\end{split}
\end{equation}
where $n = 1, 2, 3,...$ and the real parameter $\alpha>-1$ and $\beta> -1$.

 The Eq. (\ref{eq13}) can be expressed as an eigenvalue equation 
\begin{equation}
\begin{split}
\left[(x^2-1)\frac{d^2}{dx^2}+\left((\beta+\alpha+2)x-(\beta-\alpha) + \frac{2(\beta-\alpha)(x^2-1)}{(\beta+\alpha)-(\beta-\alpha)x}\right)\frac{d}{dx}\right.
\\ \left.+\left(\beta-\alpha)x+\frac{(\beta-\alpha)^2(x^2-1)}{(\beta+\alpha)-(\beta-\alpha)x}-(n-1)(n+\alpha+\beta)\right)\right]\hat{P}_n^{(\alpha,\beta)}(x)=0\\
n=1,2,3,... \,.
\end{split}
\end{equation}
This can be formally put in the form  
\begin{equation}\label{ene}
[H-(n-1)(n+\alpha+\beta)]\hat{P}_n^{(\alpha,\beta)}(x)=0
\end{equation}
where
\begin{equation}\label{y}
\begin{split}
H=(x^2-1)\frac{d^2}{dx^2}+\left((\beta+\alpha+2)x-(\beta-\alpha) + \frac{2(\beta-\alpha)(x^2-1)}{(\beta+\alpha)-(\beta-\alpha)x}\right)\frac{d}{dx}
\\ +\left(\beta-\alpha)x+\frac{(\beta-\alpha)^2(x^2-1)}{(\beta+\alpha)-(\beta-\alpha)x}\right)
\end{split}
\end{equation}
The raising and lowering operators for the $X_1$- Jacobi polynomials are
\begin{subequations}
\begin{align}
A_{\alpha,\beta}=\frac{(x-c)^2}{x-b}\frac{d}{dx}\left(\frac{1}{x-c}\right)\\
B_{\alpha,\beta}= (x^2-1)\frac{(x-b)}{(x-c)}\bigg(\frac{d}{dx}+a\bigg)-a(x^2-2bx+1)
\end{align}
\end{subequations}
with $a$, $b$ and $c$ being related to $\alpha$, $\beta$ by the following relations.
\begin{subequations}\label{x}
\begin{align}
a=\frac{\beta-\alpha}{2}\\
b=\frac{\beta+\alpha}{\beta-\alpha}\\
c=b=\frac{1}{a}
\end{align}
\end{subequations}


These raising and lowering operators hold the following relations
\begin{align}
A_{\alpha,\beta}\hat{P}_n^{(\alpha,\beta)}(x)
= (\frac{n+\alpha+\beta}{2})
\hat{P}_{n-1}^{(\alpha+1,\beta+1)}\,, \label{18}\\
B_{\alpha,\beta}\hat{P}_n^{(\alpha+1,\beta+1)} 
=2n\hat{P}_{n+1}^{(\alpha,\beta)}\,, \label{19}
\end{align}
and the Hamiltonian in equation (\ref{y}) can then  be factorized in terms 
of these two operators as
\begin{equation}
H=B_{\alpha,\beta}A_{\alpha,\beta}= A_{\alpha-1,\beta-1}B_{\alpha-1,\beta-1}-(\alpha+\beta)\\
\end{equation} 
By using Eqs. (\ref{18}) and (\ref{19}) in Eq. (\ref{ene}) we have
\begin{eqnarray}
B_{\alpha,\beta}A_{\alpha,\beta}\hat{P}_n^{(\alpha,\beta)}=E_n\hat{P}_n^{(\alpha,\beta)}\\ \nonumber
A_{\alpha,\beta}B_{\alpha,\beta}(A_{\alpha,\beta}\hat{P}_n^{(\alpha,\beta)})=E_n(A_{\alpha,\beta}\hat{P}_n^{(\alpha,\beta)})
\end{eqnarray}
with
\begin{equation}
E_n= (n-1)(n+\alpha+\beta)
\end{equation}
Thus $H^1$ and $H^2$ as defined by
\begin{subequations}
\begin{align}
H^1=B_{\alpha,\beta} A_{\alpha,\beta}\\
H^2=A_{\alpha,\beta}B_{\alpha,\beta}
\end{align}
\end{subequations}
 are supersymmetric partner Hamiltonians.

We can generalize this further and show that the pair of Hamiltonians 
$B_{\alpha+k-1,\beta+k-1} A_{\alpha+k-1,\beta+k-1}$ and  
$A_{\alpha+k-1,\beta+k-1}B_{\alpha+k-1,\beta+k-1}$  are the  supersymmetric 
partner Hamiltonians for k= 1,2,... \,.
In particular, by using Eqs. (\ref{18}) and (\ref{19}) we can write,
 \begin{equation}
 B_{\alpha+k,\beta+k} A_{\alpha+k,\beta+k}
\hat{P}_n^{(\alpha+k,\beta+k)} 
= E_n^{k}\hat{P}_n^{(\alpha+k,\beta+k)} \,, \label{a}
 \end{equation}
 and
 \begin{equation}
 A_{\alpha+k-1,\beta+k-1}B_{\alpha+k-1,\beta+k-1}
\hat{P}_{n}^{(\alpha+k,\beta+k)} 
= E_{n+1}^{k-1}\hat{P}_n^{(\alpha+k,\beta+k)}\,, \label{b}
 \end{equation} 
where
\begin{equation}
E_n^{k-1}=(n-1)(n+\alpha+\beta+2k-2)\,. 
\end{equation}
 
Above equations imply 
\begin{equation}
\begin{split}
B_{\alpha+k,\beta+k}A_{\alpha+k,\beta+k}-E_n^{k}\\
=A_{\alpha+k-1,\beta+k-1}B_{\alpha+k-1,\beta+k-1}-E_{n+1}^{k-1}\,, \label{c}
\end{split}
\end{equation}
where (k=1,2,3,...). From Eq. \eqref{c} it is clear that the supersymmetric 
pair of Hamiltonians $B_{\alpha+k-1,\beta+k-1} A_{\alpha+k-1,\beta+k-1}$ and   
$A_{\alpha+k-1,\beta+k-1}B_{\alpha+k-1,\beta+k-1}$ are in fact shape invariant 
with only a shift.\\
By using the above generalize operators we can find the $X_1$ Jacobi polynomials
\begin{equation}
  2^{(s-1)}(s-1)!\hat{P}_s^{(\alpha,\beta)}= B_{\alpha,\beta}B_{\alpha+1,\beta+1}...B_{\alpha+s-2,\beta+k-2}\hat{P}_1^{(\alpha+s-1,\beta+s-1)}
\end{equation}
where $s=2,3,4\cdots $ .
Let us consider,
\begin{equation*}
H^k= B_{\alpha+k-1,\beta+k-1}A_{\alpha+k-1,\beta+k-1}+(k-1)(k+\alpha+\beta)
\end{equation*}
 It is straight forward to see that this sequence of Hamiltonians satisfy Shape invariant property and can be written as
\begin{equation}
=H^{k+1}=H^k+(\alpha+\beta+2k)= H^k+R_k
\end{equation}
where $R_k= \left(\alpha+\beta+2k\right)$\\
So the nth eigenvalue of the  original Hamiltonian ($H^1$), as expected,
is given by
\begin{equation}
\begin{split}
E_n= \sum\limits_{k=1}^{n-1}(\alpha+\beta)+2k\\
=(n-1)(n+\alpha+\beta)\,.
\end{split}
\end{equation}

\subsection{$X_1$ Laguerre polynomials}

For $n \geq 1$ and the real parameter $k>0$ the $X_1$ Laguerre polynomials $ \hat{L}^{(k)}_n$ satisfy the differential equation 
\begin{equation}
-x\hat{L}_n^{(k)''}(x)+\frac{(x-k)}{x+k}[(x+k+1)\hat{L}_n^{(k)'}(x)]-[\frac{(x-k)}{(x+k)}+(n-1)]\hat{L}_n^{(k)}=0
\end{equation}
This differential equation can be written as the eigenvalue equation  as
\begin{equation}
\bigg (-x\frac{d^2}{dx^2}+\frac{(x-k)}{(x+k)}[(x+k+1)\frac{d}{dx}-1]
-(n-1) \bigg )\hat{L}_n^{(k)} = 0\,,
\end{equation}
which can be compactly expressed as
\begin{equation}
[H-(n-1)]\hat{L}_n^{(k)}=0\,,
\end{equation}
where
\begin{equation}
H=-x\frac{d^2}{dx^2}+\frac{(x-k)}{(x+k)}[(x+k+1)\frac{d}{dx}-1\,. \label{d}
\end{equation}
One can define the raising and the lowering operators for the $X_1$ 
Laguerre polynomials as 


\begin{subequations}
\begin{align}
A_k = -\frac{(x+k+1)^2}{x+k}\frac{d}{dx}\left(\frac{1}{x+k+1}\right)\,, \\
B_k = \frac{x(x+k)}{(x+k+1)}\bigg(\frac{d}{dx}-1\bigg)+k
\end{align}
\end{subequations}
The operators $A_k$ and $B_k$ satisfy following raising and lowering properties
\begin{subequations}
\begin{align}
A_k\hat{L}_n^{(k)}=\hat{L}_{n-1}^{(k+1)}\,, \label{e}\\ 
B_k\hat{L}_n^{(k+1)}=n\hat{L}_{n+1}^{(k)}\,. \label{f}
\end{align}
\end{subequations}
Hence the Hamiltonian in Eq. \eqref{d} can be factorized in terms 
of these two operators as 
\begin{equation}
H=B_k A_k =A_{k-1}B_{k-1}-1\,. \label{8}
\end{equation}
Using the Eqs. $\eqref{e}$ and $\eqref{f}$ we obtain 
\begin{equation}
\begin{split}
B_k A_k \hat{L}^{(k)}_n = E^{0}_{n} \hat{L}^{(k)}_n\,, \\
A_k B_k (A_k\hat{L}^{(k)}_n) =E^{0}_{n} (A_k\hat{L}^{(k)}_n)  \ \ 
\mbox{or} A_k B_k (\hat{L}^{(k+1)}_{n-1}) 
=E^{0}_{n} (\hat{L}^{(k+1)}_{n-1})\,, \end{split}
\end{equation} 
where,
\begin{equation*}
E_n^0 = (n-1)\,.
\end{equation*}
This indicates that the Hamiltonians $A_k B_k$ and $B_k A_k$  are 
supersymmetric partner Hamiltonians. Now to establish the shape invariance,
we construct the sequence of Hamiltonians $A_{k+r-1}B_{k+r-1}$ and 
$B_{k+r-1}A_{k+r-1}$ with $r=1,2,3\cdots$.
It is straightforward to check
 \begin{equation}
\begin{split}
B_{k+r-1}A_{k+r-1}\hat{L}^{(k+r-1)}_n = E^{r-1}_n \hat{L}^{(k+r-1)}\,, \\
A_{k+r-1}B_{k+r-1}(A_{k+r-1}\hat{L}^{(k+r-1)}_n) 
=E^{r-1}_n (A_{k+r}\hat{L}^{(k+r-1)}_n) \ \mbox{or} \
A_{k+r-1}B_{k+r-1}\hat{L}^{(k+r)}_{n-1} = E^{r-1}_n\hat{L}^{(k+r)}_{n-1} 
\label{g}
\end{split}
\end{equation}
with $E^{r-1}_n=(n-1)$.
 From  the recurrence relations we can write
\begin{equation}
\begin{split}
B_{k+r}A_{k+r}-(n-1)\\
=A_{k+r-1}B_{k+r-1}-n\,.
\end{split}
\end{equation}
The above relation shows that the supersymmetric partner Hamiltonians in 
Eq. $\eqref{g}$ are also shape invariant with a constant shift, 
independent of $r$. \\
We can also relate the $X_1$ Laguerre polynomials by using raising operators in the following way
\begin{equation}
   (s-1)!\hat{L}^{(k)}_{s}= B_{k} B_{k+1} B_{k+2}... B_{k+s-2}\hat{L}^{(k+s-1)}_{1}
\end{equation}
with $s=2,3,4,\cdots $.
Let us define 
\begin{equation}
H^r= B_{k+r-1}A_{k+r-1}+(r-1)\,.
\end{equation} 
We can see that the sequence of shape invariant Hamiltonians satisfy
\begin{equation}
\begin{split}
A_{k+r-1}B_{k+r-1}+(r-1)\\=B_{k+r}A_{k+r}+1+r-1\\
=B_{k+r}A_{k+r}+r\\
= H^{r+1}=H^{r}+1\,.
\end{split}
\end{equation}
Hence the $n^{th}$ eigenvalue of the Hamiltonian $H^1$ is
\begin{equation}
\begin{split}
E_n=\sum\limits_{k=1}^{n-1}1\\=(n-1)\,.
\end{split}
\end{equation} 

\section{$X_m$-  orthogonal polynomials }

The results obtained in the last section are easily generalized for the
$X_m$ EOPs as we show now.

\subsection{$X_m$- Jacobi orthogonal polynomials }

The $X_m$ Jacobi orthogonal polynomials $\hat{P}^{(\alpha,\beta)}_{n,m}(x)$ satisfy the differential equation 
\begin{equation}
\begin{split}
\hat{P''}^{(\alpha,\beta)}_{n,m}(x)+\left[(\alpha-\beta-m+1)\frac{P^{(-\alpha,\beta)}_{m-1}(x)}{P^{(-\alpha-1,\beta-1)}_m(x)}-\left(\frac{\alpha+1}{1-x}\right)+\left(\frac{\beta+1}{1+x}\right)\right]\\*\hat{P}'^{(\alpha,\beta)}_{n,m}(x)+\frac{1}{(1-x^2)}\Big[\beta(\alpha-\beta-m+1)(1-x)\frac{P^{(-\alpha,\beta)}_{m-1}(x)}{P^{(-\alpha-1,\beta-1)}_m(x)}\\+m(\alpha-\beta-m+1)+(n-m)(\alpha+\beta+n-m+1)\Big]\hat{P}^{(\alpha,\beta)}_{n,m}(x)=0
\end{split}
\end{equation}
Where $P^{(\alpha,\beta)}_{m}(x)$ are the classical Jacobi polynomials .\\
The $X_m$ Jacobi differential equation can be written as the eigenvalue  equation 
\begin{equation}
\begin{split}
 \Big[(x^2-1)\frac{d^2}{dx^2}+(x^2-1)\left((\alpha-\beta-m+1)\frac{P^{(-\alpha,\beta)}_{m-1}(x)}{P^{(-\alpha-1,\beta-1)}_m(x)}-\left(\frac{\alpha+1}{1-x}\right)+\left(\frac{\beta+1}{1+x}\right)\right)\frac{d}{dx}\\+\Big(\beta(\alpha-\beta-m+1)(x-1)\frac{P^{(-\alpha,\beta)}_{m-1}(x)}{P^{(-\alpha-1,\beta-1)}_m(x)}\\-m(\alpha-\beta-m+1)-(n-m)(\alpha+\beta+n-m+1)\Big)\Big]\hat{P}^{(\alpha,\beta)}_{n,m}(x)=0
    \end{split}
\end{equation}
This can be formally written as 
\begin{equation}
    [H-(n-m)(\alpha+\beta+n-m+1)]\hat{P}^{(\alpha,\beta)}_{n,m}(x) = 0\,,
\end{equation}
where
\begin{equation}\label{1}
\begin{split}
    H=(x^2-1)\frac{d^2}{dx^2}+(x^2-1)\left((\alpha-\beta-m+1)\frac{P^{(-\alpha,\beta)}_{m-1}(x)}{P^{(-\alpha-1,\beta-1)}_m(x)}-\left(\frac{\alpha+1}{1-x}\right)+\left(\frac{\beta+1}{1+x}\right)\right)*\frac{d}{dx}\\+\beta(\alpha-\beta-m+1)(x-1)\frac{P^{(-\alpha,\beta)}_{m-1}(x)}{P^{(-\alpha-1,\beta-1)}_m(x)}-m(\alpha-\beta-m+1)
    \end{split}
\end{equation}
The lowering and raising operators for the $Xm$ Jacobi polynomials are  
\begin{equation}
 A^m_{\alpha,\beta}=\frac{P^{(-\alpha-2,\beta)}_m(x)}
{P^{(-\alpha-1,\beta-1)}(x)}\Big[\frac{d}{dx}
-\frac{1}{2}(\beta-\alpha+m-1)\frac{P^{(-\alpha-1,\beta+1}_{m-1}(x)}
{P^{(-\alpha-2,\beta)}_m}\Big]\,,
\end{equation}
and
\begin{equation}
\begin{split}
 B^m_{\alpha,\beta}=(1-x^2)\frac{P^{(-\alpha-1,\beta-1)}_m(x)}
{P^{(-\alpha-2,\beta)}_m(x)}\Bigg[\frac{d}{dx}
-\Bigg(\frac{1}{2}(\beta-\alpha+m-1)\frac{P^{(-\alpha,\beta)}_{m-1}(x)}
{P^{(-\alpha-1,\beta-1)}_m(x)}\\+\left(\frac{\alpha+1}{1-x}\right)
-\left(\frac{\beta+1}{1+x}\right)\Bigg)\Bigg]
    \end{split}
\end{equation}

These lowering and raising operators hold the following relations 
\begin{equation}\label{2}
    A^m_{\alpha,\beta}\hat{P}^{(\alpha,\beta)}_{n,m}(x)=\frac{1}{2}(\alpha+\beta+n-m+1)\hat{P}^{(\alpha+1,\beta+1)}_{n-1,m}(x)\hspace{20mm} n\geq m
\end{equation}
and
\begin{equation}\label{3}
    B^m_{\alpha,\beta}\hat{P}^{(\alpha+1,\beta+1)}_{n,m}(x)=2(n-m+1)\hat{P}^{(\alpha,\beta)}_{n+1,m}(x)\hspace{20mm} n\geq m \,.
\end{equation}
Now by using the above recurrence relations, the Hamiltonian in (\ref{1}) 
can be factorize as
\begin{equation}
   H=B^m_{\alpha,\beta}A^m_{\alpha,\beta} 
= A^m_{\alpha-1,\beta-1}B^m_{\alpha-1,\beta-1}-(\alpha+\beta)\,.
\end{equation}
On using Eqs. (\ref{2}) and (\ref{3}),  we can write
\begin{equation}
    \begin{split}
 B^m_{\alpha,\beta}A^m_{\alpha,\beta} \hat{P}^{(\alpha,\beta)}_{n,m}(x) 
= E_{n,m}^0\hat{P}^{(\alpha,\beta)}_{n,m}(x)\,,\\
A^m_{\alpha,\beta}B^m_{\alpha,\beta}
(A^m_{\alpha,\beta}\hat{P}^{(\alpha,\beta)}_{n,m}(x)) 
= E_{n,m}^0 (A^m_{\alpha,\beta}\hat{P}^{(\alpha,\beta)}_{n,m}(x))\,, 
    \end{split}
\end{equation}
where
\begin{equation}
E_{n,m}^0=(n-m)(\alpha+\beta+n-m+1)\,.
\end{equation}
Thus the Hamiltonians
\begin{equation}
\begin{split}
    H^1=B^m_{\alpha,\beta}A^m_{\alpha,\beta}\,,\\
    H^2=A^m_{\alpha,\beta}B^m_{\alpha,\beta}\,,
\end{split}
\end{equation}
are supersymmetric partners.
 Now we construct a sequence of  Hamiltonians by introducing a parameter $k$ 
 and find that the pair of Hamiltonians 
 $B^m_{\alpha+k-1,\beta+k-1}A^m_{\alpha+k-1,\beta+k-1}$ and 
 $A^m_{\alpha+k-1,\beta+k-1}B^m_{\alpha+k-1,\beta+k-1}$ are also supersymmetric partner
 Hamiltonians for $k=1,2,3,...$ 

\begin{equation}
\begin{split}\label{3.1}
   B^m_{\alpha+k-1,\beta+k-1}A^m_{\alpha+k-1,\beta+k-1} P_{n,m}^{(\alpha+k-1,\beta+k-1)}(x)	=E^{k-1}_{n,m} P_{n,m}^{(\alpha+k-1,\beta+k-1)}(x) \\
  A^m_{\alpha+k-1,\beta+k-1} B^m_{\alpha+k-1,\beta+k-1}(A^m_{\alpha+k-1,\beta+k-1} P_{n,m}^{(\alpha+k-1,\beta+k-1)}(x)) =E^{k-1}_{n,m}(A^m_{\alpha+k-1,\beta+k-1} P_{n,m}^{(\alpha+k-1,\beta+k-1)}(x) )
   \end{split}
\end{equation}
where $k=1,2,3,...$ and
\begin{equation}
    E^{k-1}_{n,m}=(n-m)(\alpha+\beta+n-m+2k-1)
\end{equation}
We can write
\begin{equation}\label{4}
 B^m_{\alpha+k,\beta+k}A^m_{\alpha+k,\beta+k}
\hat{P}^{(\alpha+k,\beta+k)}_{n,m}(x) = 
E^k_{n,m}\hat{P}^{(\alpha+k,\beta+k)}_{n,m}(x)\,,
\end{equation}
and
\begin{equation}\label{5}
 A^m_{\alpha+k-1,\beta+k-1} B^m_{\alpha+k-1,\beta+k-1}
\hat{P}^{(\alpha+k,\beta+k)}_{n,m}(x) = 
E^{k-1}_{n+1,m}\hat{P}^{(\alpha+k,\beta+k)}_{n,m}(x)\,.
\end{equation}
From Eqs. (\ref{4}) and(\ref{5}) we can write,
\begin{equation}\label{6}
 B^m_{\alpha+k,\beta+k}A^m_{\alpha+k,\beta+k} = A^m_{\alpha+k-1,\beta+k-1} 
B^m_{\alpha+k-1,\beta+k-1}+E^k_{n,m}-E^{k-1}_{n+1,m}\,,
\end{equation}
where(k=1,2,3,...). Now from Eq. (\ref{6}) it is clear that the 
supersymmetric pairs of Hamiltonians in Eq. (\ref{3.1}) are in fact shape 
invariant with only a shift. \\
By using the above generalize operators we can write the $X_m$ Jacobi polynomials as
\begin{equation}
    2^{(s-m)}(s-m)!\hat{P}^{(\alpha,\beta)}_{s,m}(x)= B^m_{\alpha,\beta} B^m_{\alpha+1,\beta+1}.... B^m_{\alpha+s-m-1,\beta+s-m-1}\hat{P}^{(\alpha+s-m,\beta+s-m)}_{m,m}(x)
\end{equation}
Where $s =  m+1, m+2,  \cdots $ 
To see that, let us define a family of Hamiltonians 
\begin{equation}
 H^k=B^m_{\alpha+k-1,\beta+k-1} A^m_{\alpha+k-1,\beta+k-1}
+(k-1)(k+\alpha+\beta)\,.
\end{equation}
We can see that these shape invariant Hamiltonians satisfy
\begin{equation}
    \begin{split}
A^m_{\alpha+k-1,\beta+k-1} B^m_{\alpha+k-1,\beta+k-1}+(k-1)(k+\alpha+\beta)\\
= B^m_{\alpha+k,\beta+k}A^m_{\alpha+k,\beta+k}+(\alpha+\beta+2k) 
+(k-1)(k+\alpha+\beta)\,.\\
    \end{split}
\end{equation}

Hence, we can write
\begin{equation}
    H^{k+1}=H^k+R_k
\end{equation}
where$R_k=(\alpha+\beta+2k)$\\
So the $n^{th}$ eigenvalue of Hamiltonian $H^1$ is given by
\begin{equation}
\begin{split}
  E_n^m= \sum\limits_{k=1}^{n-m}(\alpha+\beta)+2k\\
  =(n-m)(\alpha+\beta+n-m+1)\,.
  \end{split}
\end{equation}
\subsection{$X_m$- Laguerre orthogonal polynomials }
For an integer $m\geq0$, the $X_m$ Laguerre orthogonal polynomials 
$\hat{L}^{k}_{n,m}(x)$ satisfy the differential equation
\begin{eqnarray}
\hat{L''}^{k}_{n,m}(x)+\frac{1}{x}\left((k+1-x)-2x\frac{L^{k}_{m-1}(-x)}
{L^{k-1}_{m}(-x)}\right)\hat{L'}^{k}_{n,m}(x) \\ \nonumber 
+\frac{1}{x}\left(n-2k\frac{L^{k}_{m-1}(-x)}{L^{k-1}_{m}(-x)}\right)
\hat{L}^{k}_{n,m}(x)= 0\,,
\end{eqnarray}
where $L^{k}_{m}$ are the classical Laguerre polynomials.

The above differential equation  can be written in the form of eigenvalue 
equation as
\begin{equation}
    \begin{split}
\Bigg[-x\frac{d^2}{dx^2}+\left((x-k-1)+2x\frac{L^k_{m-1}(-x)}{L^{k-1}_{m}(-x)}
\right)\frac{d}{dx}\\
+\left(2k\frac{L^k_{m-1}(-x)}{L^{k-1}_{m}(-x)}-n\right)\Bigg] \hat{L}^k_{n,m}(x) = 0\,.
    \end{split}
\end{equation}
This can be formally put in the form 
\begin{equation}
[H-(n-m)]\hat{L}^k_{n,m}(x) = 0\,,
\end{equation}
where
\begin{equation}\label{7.1}
H=-x\frac{d^2}{dx^2}+\left((x-k-1)+2x\frac{L^k_{m-1}(-x)}{L^{k-1}_{m}(-x)}\right)\frac{d}{dx}\\
    +\left(2k\frac{L^k_{m-1}(-x)}{L^{k-1}_{m}(-x)}-m\right)\,.
\end{equation}
The lowering and raising operators for the $X_m$ Laguerre orthogonal 
polynomials are
\begin{equation}
A^{m}_{k}=-\frac{L^{k}_{m}(-x)}{L^{k-1}_{m}(-x)}\left(\frac{d}{dx}
-\frac{L^{k+1}_{m-1}(-x)}{L^{k}_{m}(-x)}\right)\,,
\end{equation}
and
\begin{equation}
B^{m}_{k}=\frac{L^{k-1}_{m}(-x)}{L^{k}_{m}(-x)}\left(x\frac{d}{dx}
+(1+k)\right)-x\,.
\end{equation}
These operators hold the following relations\\
\begin{equation}\label{8}
A^{m}_{k}\hat{L}^{k}_{n,m}(x)=\hat{L}^{k+1}_{n-1,m}(x)
\end{equation}
and 
\begin{equation}\label{9}
 B^{m}_{k}\hat{L}^{k+1}_{n,m}(x)=(n-m+1) \hat{L}^{k}_{n+1,m}(x)
\end{equation}
From the above two recurrence relations we can write the Hamiltonian in Eq.
 (\ref{7.1}) in a factorize form as 
\begin{equation}
   H=B^{m}_{k} A^{m}_{k} =A^{m}_{k-1} B^{m}_{k-1}-1\,.
\end{equation}
On using Eqs. (\ref{8}) and (\ref{9}) we can write 
\begin{equation}
    \begin{split}
 B^{m}_{k}A^{m}_{k}\hat{L}^{k}_{n,m}(x) =E^0_{n,m}\hat{L}^{k}_{n,m}(x)\\
 A^{m}_{k} B^{m}_{k} (A^m_k \hat{L}^k_{n,m}(x)) 
= E^0_{n,m}( A^m_k\hat{L}^k_{n,m}(x))\,,
    \end{split}
\end{equation}
where $E^0_{n,m}=(n-m)$.     \\
The above conditions show that the Hamiltonians $ A^{m}_{k} B^{m}_{k}$ and 
$B^{m}_{k} A^{m}_{k} $  are supersymmetric partner Hamiltonians .\\

Now we introduce a parameter $r$ and find that the pairs of Hamiltonians 
$A^{m}_{k+r-1} B^{m}_{k+r-1}$ and $B^{m}_{k+r-1} A^{m}_{k+r-1}$ are supersymmetric 
partners, for $r=1,2,3 \cdots ,$ by satisfying the conditions 
\begin{equation}
    \begin{split}\label{10}
B^{m}_{k+r-1} A^{m}_{k+r-1}\hat{L}^{k+r-1}_{n,m}(x) = E^{r-1}_{n,m}\hat{L}^{k+r-1}_{n,m}(x) \\
 A^{m}_{k+r-1} B^{m}_{k+r-1} (A^m_{k+r-1}\hat{L}^{k+r-1}_{n,m}(x)) 
    =E^{r-1}_{n,m} (A^m_{k+r-1}\hat{L}^{k+r-1}_{n,m}(x))\,,    \end{split}
\end{equation}
where $E^{r-1}_{n,m}=(n-m).$\\  We see that the eigenvalues of the Hamiltonians are independent of r. 
So all $n$ eigenstates have the degenerate eingenvalues.\\
From the recurrence relation we can write 
\begin{equation}
    \begin{split}
B^{m}_{k+r}A^{m}_{k+r}-(n-m)\\
=A^{m}_{k+r-1}B^{m}_{k+r-1}-(n-m+1)\,.
    \end{split}
\end{equation}
The above relation shows that the supersymmetric pair of Hamiltonians in 
Eq. (\ref{10}) are in fact shape invariant with a constant shift.\\

    We can also relate the $X_m$ Laguerre polynomials by using raising operators in the following way
\begin{equation}
    (s-m)!\hat{L}^{(k)}_{s,m}= B_{k}^m B_{k+1}^m B_{k+2}^m... B_{k+s-m-1}^m\hat{L}^{(k+s-m)}_{m,m}
\end{equation}
where $s= m, m+1, \cdots $.
Let us define 
\begin{equation}
 H^r=B^{m}_{k+r-1} A^{m}_{k+r-1}+(r-1)
\end{equation}
we can see that the sequence of shape invariant Hamiltonian satisfy
\begin{equation}
\begin{split}
A^{m}_{k+r-1} B^{m}_{k+r-1}+(r-1)\\
=B^{m}_{k+r} A^{m}_{k+r}+r\,,\\
   \end{split}
\end{equation}
or in other words $H^{r+1} = H^{r} +1$
 so that the n'th eigen value of the Hamiltonian is
 \begin{equation}
 E_{n}^{m}=\sum\limits_{k=1}^{n-m}1\\=(n-m)\,.
 \end{equation}
 
\section{Conclusion}

Exceptional Laguerre and exceptional Jacobi polynomials are used  extensively in the rational extension of many quantum mechanical problems and hence worth investigating them in details. In this paper we have described a new way of looking at the  differential equations for these EOPs and their solutions 
in the framework of supersymmetry and shape
invariance.  Supersymmetry and Shape invariance of the ``Hamiltonians " corresponding to $X_1$-Jacobi  and $X_1$ -Laguerre polynomials have explicitly been shown in this work. The underlying shape invariance symmetry is responsible for the solubility of  differential equations associated with these EOPs. Further these results are then generalised for $X_m$- Jacobi  and $X_m$ Laguerre polynomials with $m=1,2,3 \cdots$. We would like to add that classical Legendra   and  hypergeometric polynomials have been analysed in this fashion in \cite{bd,dk}. In the present work we have extended their work for the EOPs. In reference \cite{maj} some master conditions are
provided to check this for classical polynomials. It will be interesting to investigate EOPs along the line of \cite{maj} to obtain similar master condition for EOPs.

{\bf Acknowledgments:} BPM  acknowledges the support from MATRIX project (Grant No. MTR/2018/000611), SERB, DST Govt. of India and the Research Grant for Faculty under IoE Scheme  (number 6031) of BHU. SY acknowledges CSIR-HRDG, New Delhi, India for JRF fellowship (File No: 09/013 (0918)/2019-EMR-I)\\

\end{document}